\documentclass[conference]{IEEEtran}
\IEEEoverridecommandlockouts

\usepackage{cite}
\usepackage{amsmath,amssymb,amsfonts}
\usepackage{algorithmic}
\usepackage{graphicx}
\usepackage{textcomp}
\usepackage{xcolor}
\usepackage{svg}
\usepackage[caption=false,font=footnotesize]{subfig}
\makeatletter
 
\renewcommand{\p@subfigure}{\thefigure} 
 
\makeatother
\usepackage{booktabs}
\usepackage{multirow}
\usepackage{enumitem}
\usepackage[colorlinks=true, linkcolor=blue, citecolor=blue, urlcolor=blue]{hyperref}
\usepackage{aliascnt}

\def\BibTeX{{\rm B\kern-.05em{\sc i\kern-.025em b}\kern-.08em
    T\kern-.1667em\lower.7ex\hbox{E}\kern-.125emX}}

\begin{document}

\title{Dynamic Stability Assessment of Grid-Connected Data Centers Powered by Small Modular Reactors}


\author{Sobhan~Badakhshan,~\IEEEmembership{Graduate Student Member,~IEEE},  Roshni Anna Jacob,~\IEEEmembership{Member,~IEEE}, \\Ali Mahboub Rad,~\IEEEmembership{Graduate Student Member,~IEEE}, Chao Pan, Yaoyu Li, and Jie Zhang,~\IEEEmembership{Senior Member,~IEEE}

}


\maketitle

\begin{abstract} The accelerating growth of computational demand in modern data centers has further heightened the need for power infrastructures that are highly reliable, environmentally sustainable, and capable of supporting grid stability. Small Modular Reactors (SMRs) as a clean source of energy are particularly attractive for next-generation hyperscale data centers with significant electrical and cooling demands. This paper presents a comprehensive dynamic modeling and stability analysis of a grid-connected Integrated Energy System (IES) designed for data center applications. The proposed IES integrates an SMR and a battery energy storage system to jointly supply electricity for computational and cooling load while providing stability support to the main grid. A coupled computational-thermal load model is developed to capture the real-time power demand of the data center, incorporating CPU utilization, cooling efficiency, and ambient temperature effects. The integrated SMR-powered data center model is implemented in PSS®E and tested on the IEEE 118-bus system under various fault scenarios. Simulation results demonstrate that the IES substantially enhances voltage and frequency stability compared to a conventionally grid-connected data center, minimizing disturbance-induced deviations and improving post-fault recovery. 
\end{abstract}


\section{Introduction} 
The increasing energy demands of modern data centers, coupled with the need for high reliability and low carbon footprints, have motivated the integration of advanced energy systems. The rapid advancement of artificial intelligence (AI), particularly the proliferation of large language models (LLMs) and generative AI applications, has led to a significant surge in hyperscale data center energy consumption. These facilities typically operate 24/7, with power demands that can fluctuate rapidly due to the dynamic nature of AI workloads. Such characteristics pose unique challenges to power grid stability and reliability. 

Global electricity use by data centers is projected to more than double by 2030, reaching about 945 terawatt-hours (TWh), mainly driven by AI workloads. From 2024 to 2030, electricity consumption in data centers is expected to increase by 15\% annually, over four times faster than the growth of electricity use in other sectors \cite{IEA2025}. In the United States, data centers alone could account for up to 12\% of total electricity consumption by 2030 \cite{Shehabi2024}. This unprecedented growth is straining existing grid infrastructure, which was not designed to accommodate such large, concentrated, and variable loads.
In the context of modern energy management, Integrated Energy Systems (IES) have emerged as a key approach for optimizing resource utilization and advancing sustainability \cite{11232457}. These systems interconnect diverse energy sources, including renewable generation, advanced nuclear reactors, low-carbon fuels, and energy storage technologies, to efficiently meet the demands of both local and regional consumers \cite{JIA2024120328}. IES can combine multiple energy carriers to supply electricity, heat, transportation, and other energy services. By leveraging a variety of resources and integrating different subsystems, IES provides effective solutions to the pressing challenges of modern energy systems, including resilience, reliability, and environmental sustainability \cite{ARENT202147}. 
Recent research has focused extensively on the optimal design, operation, and control of IES \cite{Mengshu}, highlighting their potential to improve efficiency, cost-effectiveness, and overall system reliability when integrated with the main power grid \cite{Tao9732191}. 


Deploying an IES at the data center level offers several advantages for both local energy supply and the broader grid. By generating and managing both electrical and thermal energy on-site, the IES enhances the overall efficiency, sustainability, and reliability of data center operations \cite{WANG2023126585,GUO2021117474}. However, modern data centers supporting AI applications such as large language models (LLMs) experience extremely fast-varying and unpredictable workloads, leading to significant power demand fluctuations. Traditional energy models and static provisioning approaches, originally designed for more stable workloads, are often inadequate to capture these dynamics, motivating the need for advanced IES-based frameworks capable of dynamically responding to rapid load variations while maintaining efficiency and grid compatibility. The small modular reactor (SMR) provides a stable, carbon-free power source with high availability, while the battery energy storage system enables rapid response to load variations and transient events \cite{11028096}. Furthermore, the thermal energy produced by the SMR can be harnessed for heating or cooling within the data center’s HVAC system, reducing energy waste and improving overall system utilization. This integrated approach decreases dependence on the main grid, mitigates transmission losses, and strengthens local energy autonomy, making the SMR-based IES a promising solution for achieving sustainable and resilient data center power infrastructure. The data center load is modeled as a combination of computational demand, characterized by the number of CPU cores, total CPU capacity, and utilization, and thermal load associated with the cooling infrastructure.

In this paper, we present a comprehensive study on the integration of SMR and Battery Energy Storage System (BESS) within a grid-connected IES for data centers. The novelty of this work lies in modeling the SMR as an energy source, that simultaneously serves the IT load and cooling systems, while coordinating it with fast-acting battery support to ensure robust frequency regulation. The data center load is represented using actual power demand values that reflect real-time variations in server utilization and cooling requirements. For reliable operation and stability analysis, pre-fault voltage and frequency conditions are obtained from steady-state power flow analysis. Then, high-resolution transient simulations are performed to evaluate the system's response to realistic disturbances such as faults, line trips, and sudden load changes. The SMR is modeled using a modified GGOV1 governor with setpoint-based droop control, providing adaptive and safe frequency response under varying electrical and thermal loads, and the BESS delivers rapid transient support through a proportional-integral (PI) control strategy. 

The rest of this article is organized as follows: Section \ref{sec:dymmodeling} discusses the dynamic modeling of SMR-based IES and the data center. In Section \ref{Meth}, we present the methodology for stability analysis of the grid connected with an IES datacenter. Results from the case study are presented in Section \ref{Result}, and conclusions are drawn in Section \ref{conc}.

\section{Dynamic Modeling of SMR-based Data Center}\label{sec:dymmodeling}

\subsection{IES Dynamic Modeling}
This section presents the dynamic modeling framework for the IES, which couples the SMR and BESS to supply the data center. The model is developed wit goal of capturing the dynamic behavior and frequency response capability of the IES under varying operating conditions and disturbances. 
The SMR unit is modeled using a modified GE General Governor (GGOV1) framework to accurately capture its frequency control characteristics. GGOV1 is chosen for its ability to represent both isochronous and droop control modes, along with integrated load limiter functionality critical for ensuring reactor safety and maintaining thermal constraints \cite{ggo}. The governor architecture includes rate limits on valve actuation, providing physically realistic turbine dynamics \cite{10220527}.

In this implementation, the SMR employs setpoint-based control, enabling frequency-dependent power adjustments through coordinated modulation of steam valve positions. A load limiter enforces thermal safety by constraining the permissible rate of power change during transients. \textcolor{black}{Specifically, the load limiter restricts rapid reactor power changes that could cause excessive fuel or coolant temperature excursions. By enforcing predefined ramp-rate limits based on thermal–hydraulic design constraints, it overrides governor commands when necessary to ensure safe operation during load-following and transient events.}


The mechanical power output of the turbine is expressed as
\begin{equation}
P_\text{mech} = \eta_T \left( \Delta H^\text{HP}\, \dot{m}^\text{HP}_{\text{cs}}
+ \Delta H^\text{LP}\, \dot{m}^\text{LP}_{\text{cs}} \right),
\label{pmech}
\end{equation}
where $\eta_T$ denotes the turbine efficiency, $\Delta H^\text{HP}$ and $\Delta H^\text{LP}$ are the enthalpy differences across the high- and low-pressure turbine stages, respectively, and $\dot{m}^\text{HP}_{\text{cs}}$ and $\dot{m}^\text{LP}_{\text{cs}}$ represent the corresponding controlled-steam mass flow rates. SMR setpoints are calibrated using Integral Pressurized Water Reactor (iPWR) simulator data to ensure consistency between the model and realistic pressure-temperature dynamics \cite{thermal}.

Frequency regulation is achieved through a droop-based control law, linking system frequency deviations to mechanical power adjustments:
\begin{equation}
\Delta P_\text{mech} = -\frac{\Delta f}{m(\dot{Q}, P_e)},
\end{equation}
where $\Delta f$ is the frequency deviation and $m(\dot{Q}, P_e)$ is the droop coefficient dependent on the thermal load $\dot{Q}$ and electrical power output $P_e$. The negative sign reflects the inverse relationship between frequency deviation and power correction.

The droop coefficient is adjusted based on electrical and thermal loading:
\begin{equation}
m(\dot{Q}, P_e) = m_\mathrm{min} + 
\frac{P_e + \dot{Q}}{P_\mathrm{max} + \dot{Q}_\mathrm{max}}
(m_\mathrm{max} - m_\mathrm{min}),
\end{equation}
where $m_\mathrm{min}$ and $m_\mathrm{max}$ are the minimum and maximum droop values, respectively. At low load, greater power margin allows a stronger frequency response (smaller $m$), whereas at high load the reduced margin weakens the response (larger $m$). This ensures safe and effective participation in primary frequency control while respecting thermal constraints.
The frequency control loop modulates mechanical power based on $\Delta f$, increasing power under under-frequency events and reducing it under over-frequency conditions, thus enabling coordinated participation of the SMR in primary frequency regulation. The model is implemented as a user-defined module in PSS®E, facilitating transient simulations under step disturbances, load fluctuations, and fault events.

To address the relatively slow thermal response of the SMR, a BESS is integrated to provide fast frequency regulation and power balancing. The BESS compensates for transient power deficits almost instantaneously, supporting system stability before the SMR governor reaches steady state. Its active power output is governed by a PI controller:
\begin{equation}
P^B(t) = k_p \Delta f(t) + k_i \int_0^t \Delta f(\tau) , d\tau,
\label{pBESS}
\end{equation}
where $k_p$ and $k_i$ are the proportional and integral gains. The PI control law ensures rapid frequency response while eliminating steady-state errors, complementing the SMR’s slower dynamics and enhancing overall system robustness.

The BESS dynamics are implemented using the REECC (Renewable Energy Electrical Controller C), REGCA (Renewable Energy Generator/Converter A), and REPCA (Renewable Plant Controller A) for system-level control available in PSS®E.

The integrated dynamic IES model couples the SMR’s thermal-electrical behavior with the BESS’s fast electrical response to achieve multi-timescale frequency control. This coordination enables reliable and resilient data center operation by combining rapid frequency regulation with steady state power generation. Simulations in PSS®E confirm that the SMR and BESS work complementarily to maintain stable power and frequency under step disturbances and fault conditions.

\subsection{Data Center Load Modeling}
This study models the data center load as the sum of IT equipment power consumption and thermal load, accounting for both energy use and cooling demands. The IT load constitutes the major portion of a data center’s total electrical demand and is primarily responsible for its temporal fluctuations. The IT load profile considered here is derived from the publicly available Google Cluster workload traces~\cite{reiss2012heterogeneity}, which provide records of task-level CPU utilization and machine events collected from a large-scale production data center. The raw usage data are processed to obtain a 5-minute resolution, cluster-level CPU utilization trace. Each task’s active interval, defined by its start and end times, is mapped onto fixed 5-minute bins, and its CPU consumption is accumulated in proportion to the overlap duration within each bin. The results are then aggregated across all machines, while the total available computing capacity is dynamically estimated using the machine-event logs that record node additions, removals, and capacity updates throughout the trace. This step results in a normalized utilization curve that represents the aggregate fraction of active computing resources over time. The utilization trace is then translated into an equivalent IT power demand using an affine power model consistent with empirical datacenter measurements.
\begin{equation}
P_{\text{IT}}(t) = P_{\text{idle}} + \big(P_{\text{max}} - P_{\text{idle}}\big) u_{\text{cpu}}(t),
\label{eq:itload}
\end{equation}
where $P_{\text{IT}}(t)$ is the instantaneous IT power demand, $u_{cpu}(t)$ is the normalized CPU utilization, $P_{max}$ is the peak IT capacity, and $P_{\text{idle}}=0.5P_{max}$ represents the idle-power fraction typical of hyperscale data centers. This captures the quasi-linear relationship between server power and CPU utilization while preserving the temporal variability observed in real computing workloads.

\begin{figure}[htbp!]
  \centering \includegraphics[width=0.9\linewidth]{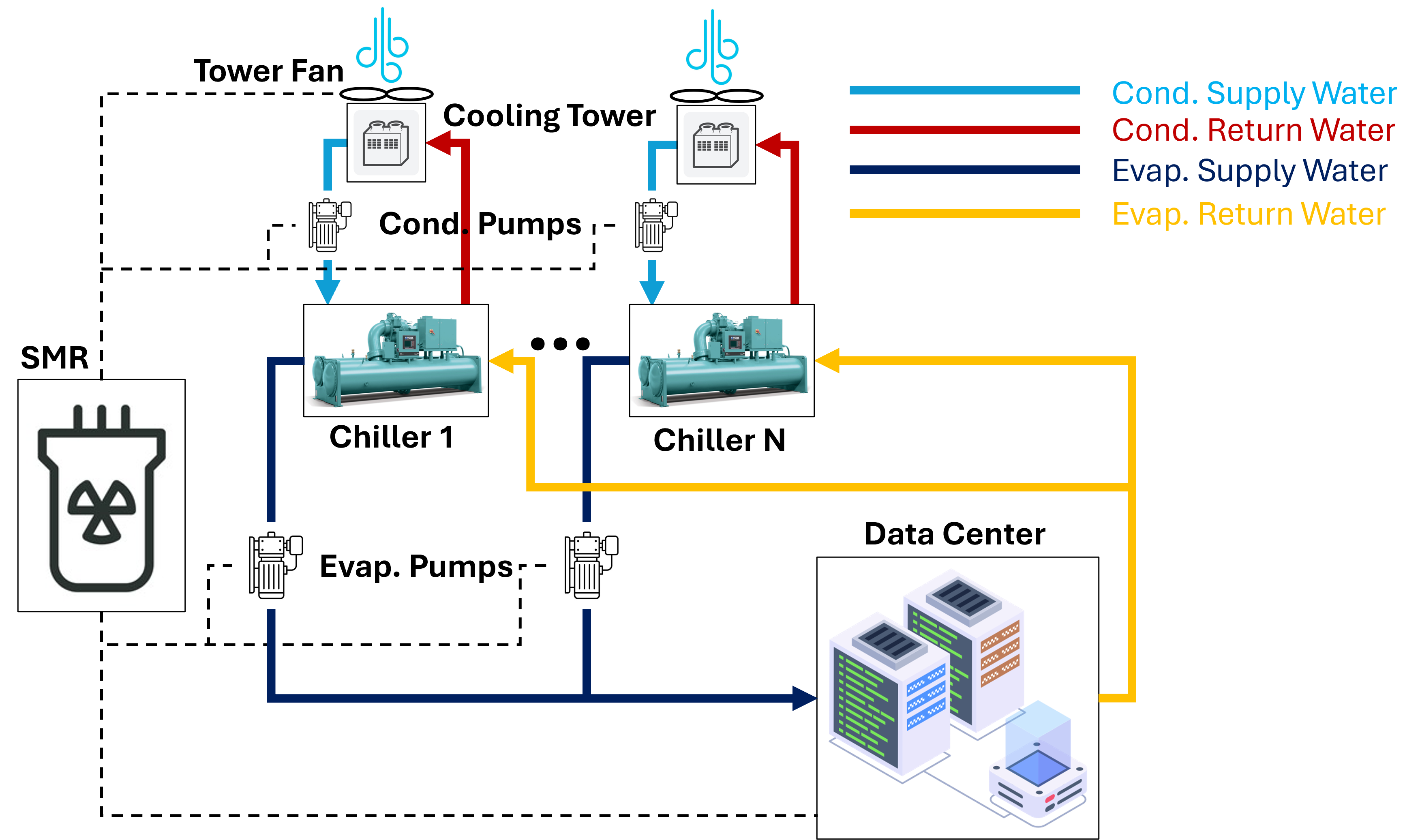}
  \caption{Schematic of the chiller cooling system for the data center} 
  \label{coolingsystem}
\end{figure}

Data centers also require substantial cooling to maintain safe operating temperatures. As illustrated in Fig. \ref{coolingsystem}, a chiller bank provides the necessary cooling to regulate the data center environment under varying operating conditions. The cooling system's power consumption includes the tower fans, condenser pumps, evaporator pumps, and chiller compressors. A cooling coil is used with a chilled water loop and a recirculating air loop in the data center, where we consider that the chilled water can always remove the thermal load from the data center. Therefore, the return-chilled water condition is associated with the cooling demand from the data center.

For each chiller unit in the chiller bank, the total electrical power ($P_{\text{ch}}$) consumption is the sum of a tower fan power ($P_{\text{tf}}$), condenser pump power ($P_{\text{cd}}$), evaporator pump power ($P_{\text{ev}}$), and compressor power ($P_{\text{cmp}}$):
\begin{subequations}
\begin{equation}
P_{\text{ch}}(t) = P_{\text{tf}}(t) + P_{\text{cd}}(t) + P_{\text{ev}}(t) + P_{\text{cmp}}(t),
\label{eq:chillerpower}
\end{equation}
\begin{equation}
P_{\text{tf}}(t) = {\alpha}_{1} \dot m_{\text{tf}}(t) + {\alpha}_{2}^2 \dot m_{\text{tf}}(t) + {\alpha}_{3}^3 \dot m_{\text{tf}}(t) + {\alpha}_{4},
\label{eq:towerfanpower}
\end{equation}
\begin{equation}
P_{\text{cd}}(t) = {\beta}_{1} \dot m_{\text{cd}}(t) + {\beta}_{2}^2 \dot m_{\text{cd}}(t) + {\beta}_{3}^3 \dot m_{\text{cd}}(t) + {\beta}_{4},
\label{eq:condpumppower}
\end{equation}
\begin{equation}
P_{\text{ev}}(t) = {\gamma}_{1} \dot m_{\text{ev}}(t) + {\gamma}_{2}^2 \dot m_{\text{ev}}(t) + {\gamma}_{3}^3 \dot m_{\text{ev}}(t) + {\gamma}_{4},
\label{eq:evappumppower}
\end{equation}
\begin{equation}
\begin{split}
P_{\text{cmp}}(t) = f_{cmp}(T_{rw}(t),T_{amb}(t),\phi _{amb}(t),\\ \dot m_{\text{tf}}(t),\dot m_{\text{cd}}(t),\dot m_{\text{ev}}(t)).
\end{split}
\label{eq:evappumppower}
\end{equation}
\end{subequations}
Here, $P_{\text{tf}}$, $P_{\text{cd}}$, and $P_{\text{ev}}$ are modeled as cubic-order functions of their respective mass flow rates. The compressor power $P_{\text{cmp}}$ is identified as a black-box model with inputs and disturbances data, where $T_{rw}$ represents the return-chilled water temperature, $T_{amb}$ and $\phi _{amb}$ denote the ambient temperature and relative humidity, respectively. Due to the variable IT load, the number of active chillers in the bank changes dynamically to meet the cooling demand. Accordingly, the total thermal load from the chiller bank is formulated as:
\begin{equation}
P_{\text{thermal}}(t) = n_{ch}(t) P_{\text{ch}}(t).
\label{eq:totalthermalpower}
\end{equation}
where $n_{ch}$ represents the operating chiller numbers. The above chiller cooling system is developed based on previous work in \cite{Pan2024chiller} with Dymola \cite{Dymola2019} and TIL library \cite{TILlibrary}.

\section{Methodology}\label{Meth}
To evaluate the impact of an IES-powered data center on grid stability, a two-step approach is adopted, combining steady-state power flow analysis and transient disturbance simulation. For each 5-minute time step, the steady-state power flow of the system is calculated to determine pre-fault voltage and frequency conditions at the point of interconnection between the data center (or IES) and the main grid. The 5-minute resolution is chosen because it captures both the natural variations in data center load due to changing computational demand and cooling requirements, and allows sufficient temporal granularity for observing system dynamics, without imposing excessive computational burden. This establishes a reliable baseline operational state, capturing the normal operating conditions of the data center load, including both computational and thermal demands. The steady-state analysis provides a clear understanding of how the system behaves under typical load variations and ensures that any observed dynamic response during transient events can be accurately attributed to disturbances rather than normal operational fluctuations.

Following the steady-state analysis, transient disturbances are introduced in the main grid to simulate realistic contingencies such as short circuits, line trips, or sudden load changes. The dynamic response of the system is recorded, with particular focus on voltage and frequency at the point of interconnection between the IES datacenter and the grid. By analyzing these transient responses, the ability of the system to maintain stable operation under fault conditions can be assessed. This methodology is especially well-suited for consider large data center loads as they induce or may be highly sensitive to voltage and frequency fluctuations; even small deviations can disrupt operations or trigger protective mechanisms. Capturing the transient response at high temporal resolution allows for a detailed evaluation of how the IES, with its combination of SMR baseload generation and fast-acting battery support, mitigates these disturbances.

Finally, the methodology compares two configurations: (i) a data center connected directly to the grid, and (ii) a data center supplied through the SMR-based IES and interconnected with the grid. By examining the differences in voltage and frequency profiles under identical disturbances, this approach can be used to investigate the stability benefits provided by the IES. This method is effective for data center modeling because it captures both the time-varying load characteristics and the dynamic interaction between the grid and local energy resources, providing a comprehensive assessment of operational resilience under realistic operating conditions.

\section{Results and Discussion}\label{Result}
\begin{figure}[htb!]
\vspace{-10pt}
  \centering \includegraphics[width=0.8\linewidth]{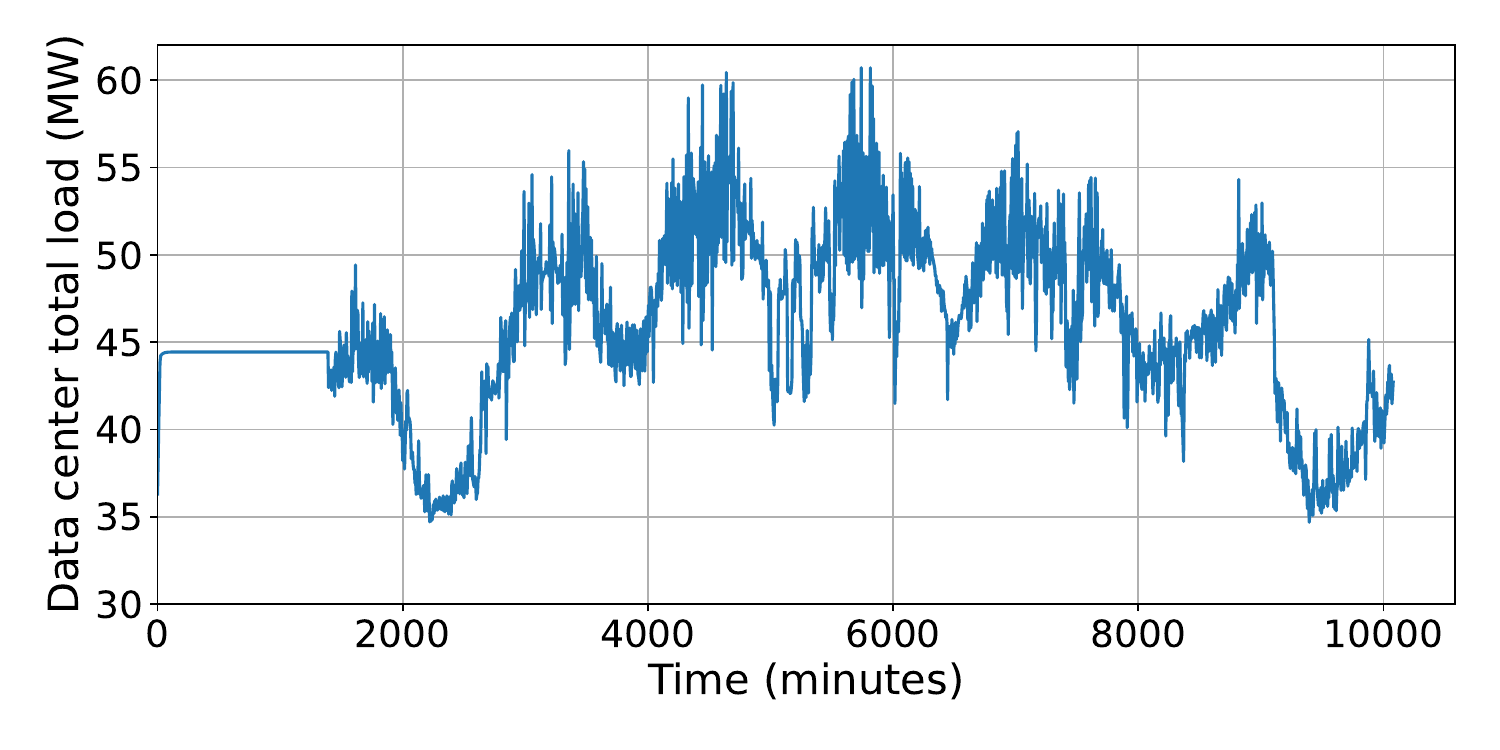}
    \vspace{-15pt}
\caption{Weekly data center total load profile over time} 
  \label{dcload}
\end{figure}

The proposed SMR-based IES for the data center was simulated using PSS\textregistered{}E and interconnected to a standard IEEE 118-bus system. The data center load profile used in the simulations spans a full week, with a 5-minute time resolution, capturing both computational and cooling load variations. For each load level, an AC power flow was performed to obtain the pre-fault operating conditions, including voltage and frequency at the point of interconnection. The maximum observed total data center load during the simulation period reached 60 MW, and the full load profile (including both IT and cooling load) is illustrated in Fig.~\ref{dcload}.

\begin{figure}[htb!]
  \centering            
  \includegraphics[width=0.85\linewidth]{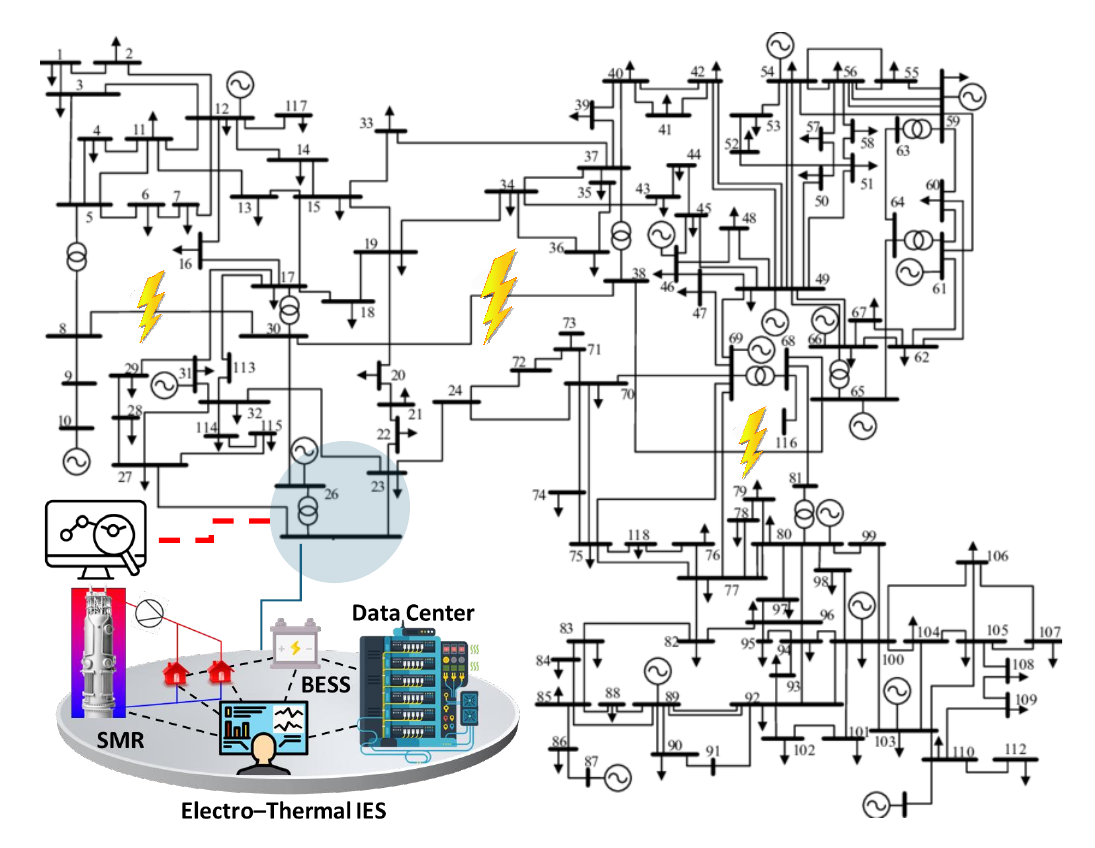}
  \caption{IES with an SMR and BESS, supplying IT and cooling load of the data center connected within the IEEE 118-bus transmission grid for stability analysis considering random fault.} 
  \label{Fig:IEStoGrid}
  \vspace{-1em}
\end{figure}

As shown in Fig.~\ref{Fig:IEStoGrid}, an SMR-powered data center is modeled and connected to the IEEE 118-bus system at bus-25.

\begin{figure}[!thb]
\centering
\vspace{-15pt}
\subfloat[Voltage Response]{%
    \includegraphics[width=2.6in]{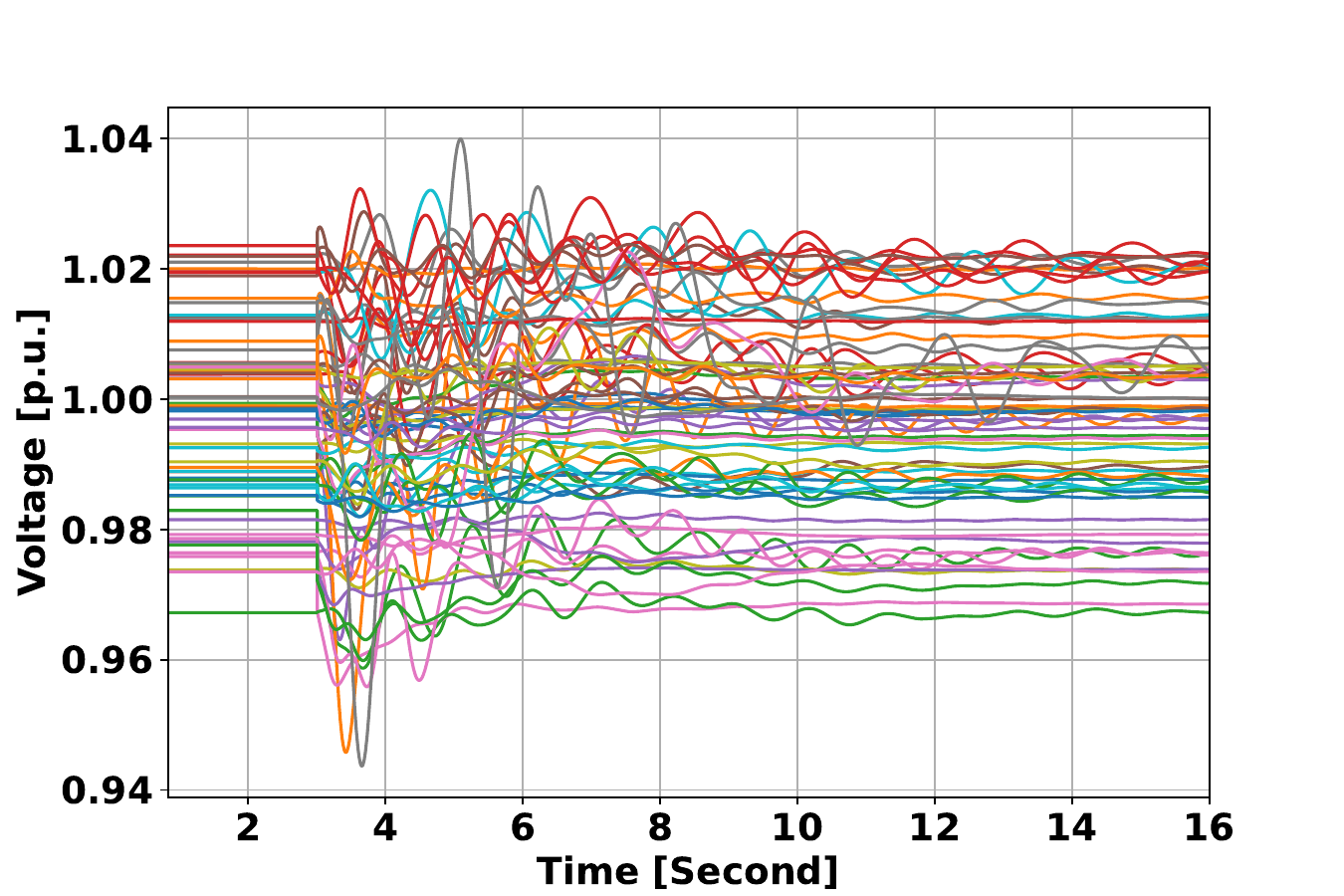}
    \label{fig:NoIes_voltage} 
}
\hspace{0.1in}
  \vspace{-10pt}

\subfloat[Dynamic Frequency Deviation]{%
    \includegraphics[width=2.6in]{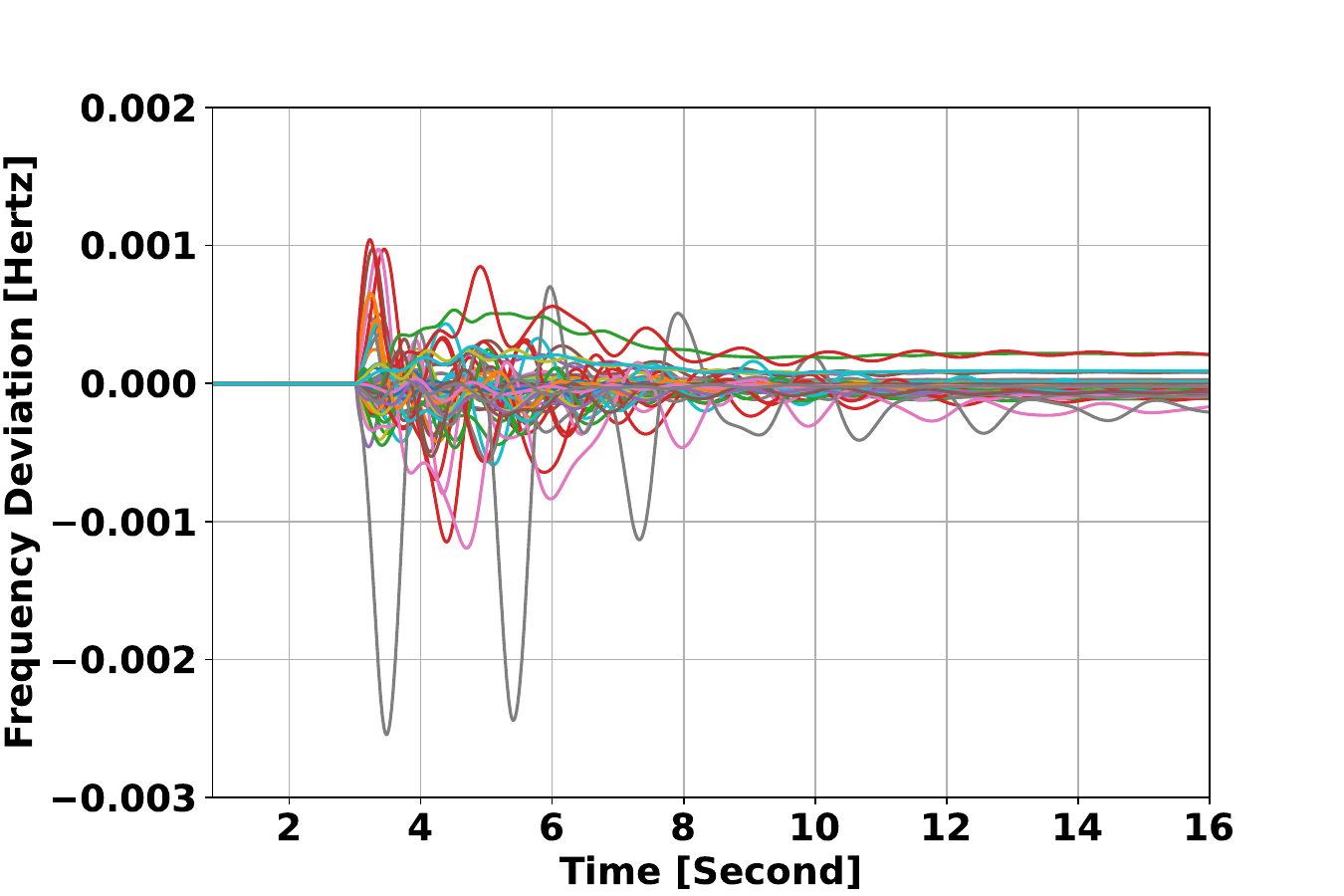}
    \label{fig:NoIes_frequency}
}
\caption{Voltage and frequency response of the data center load connected to the grid without integrated energy system.}
\label{NoIes}
\end{figure}

To assess system resilience, random faults were introduced in the main grid, including line outages, bus faults, and generator contingencies. The voltage and frequency at the interconnection point of a conventional data center without IES under these disturbances are presented in Fig.~\ref{NoIes}. The faults were applied at time 3.0, and the pre-fault voltage and frequency conditions at each time step were derived from the steady-state power flow analysis, reflecting the variations due to the changing data center load.

\begin{figure}[!thb]
\centering
\vspace{-15pt}
\subfloat[Voltage Response]{%
    \includegraphics[width=2.6in]{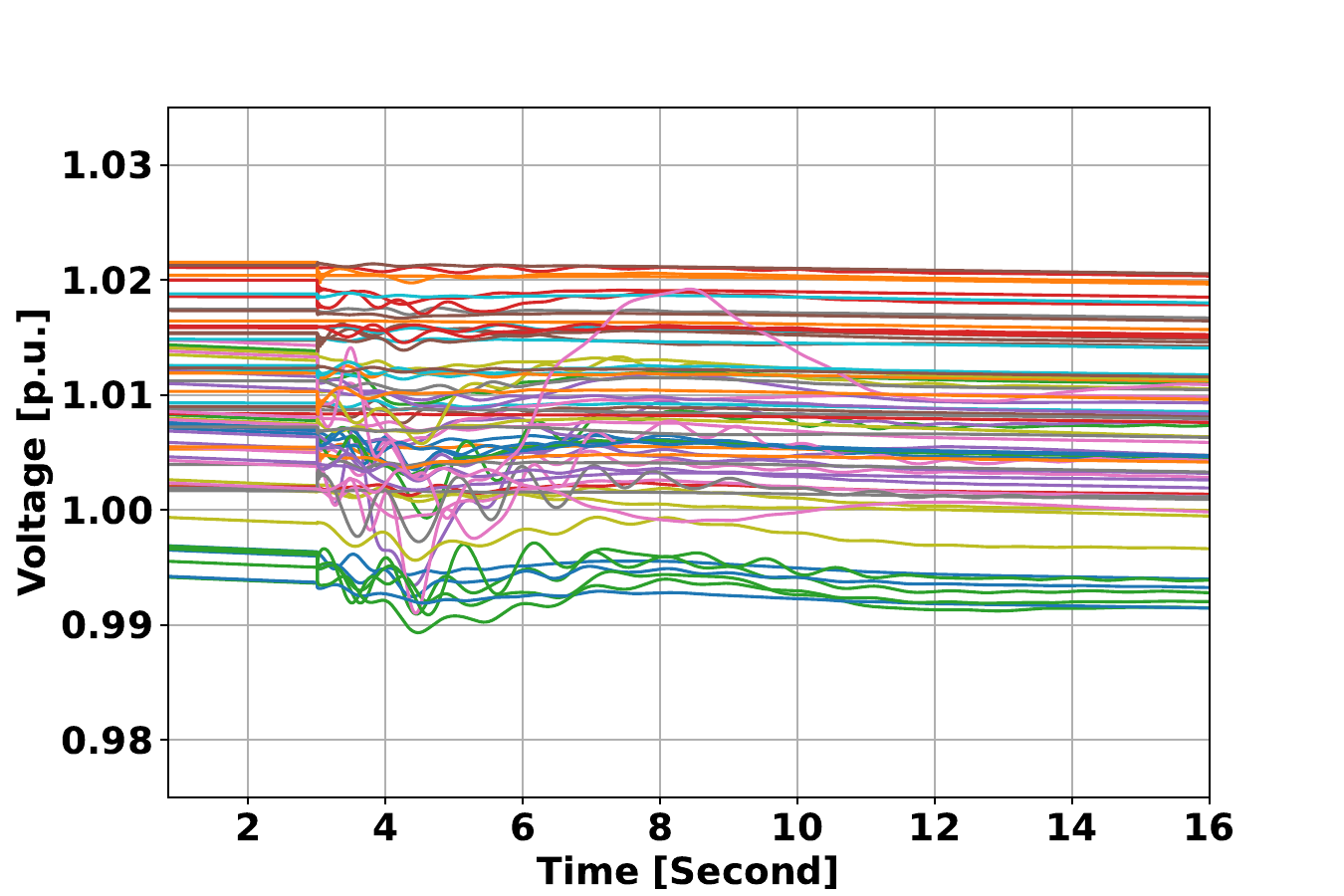}
    \label{fig:NoIes_voltage}
}
\hspace{0.1in}
  \vspace{-10pt}

\subfloat[Dynamic Frequency Deviation]{%
    \includegraphics[width=2.6in]{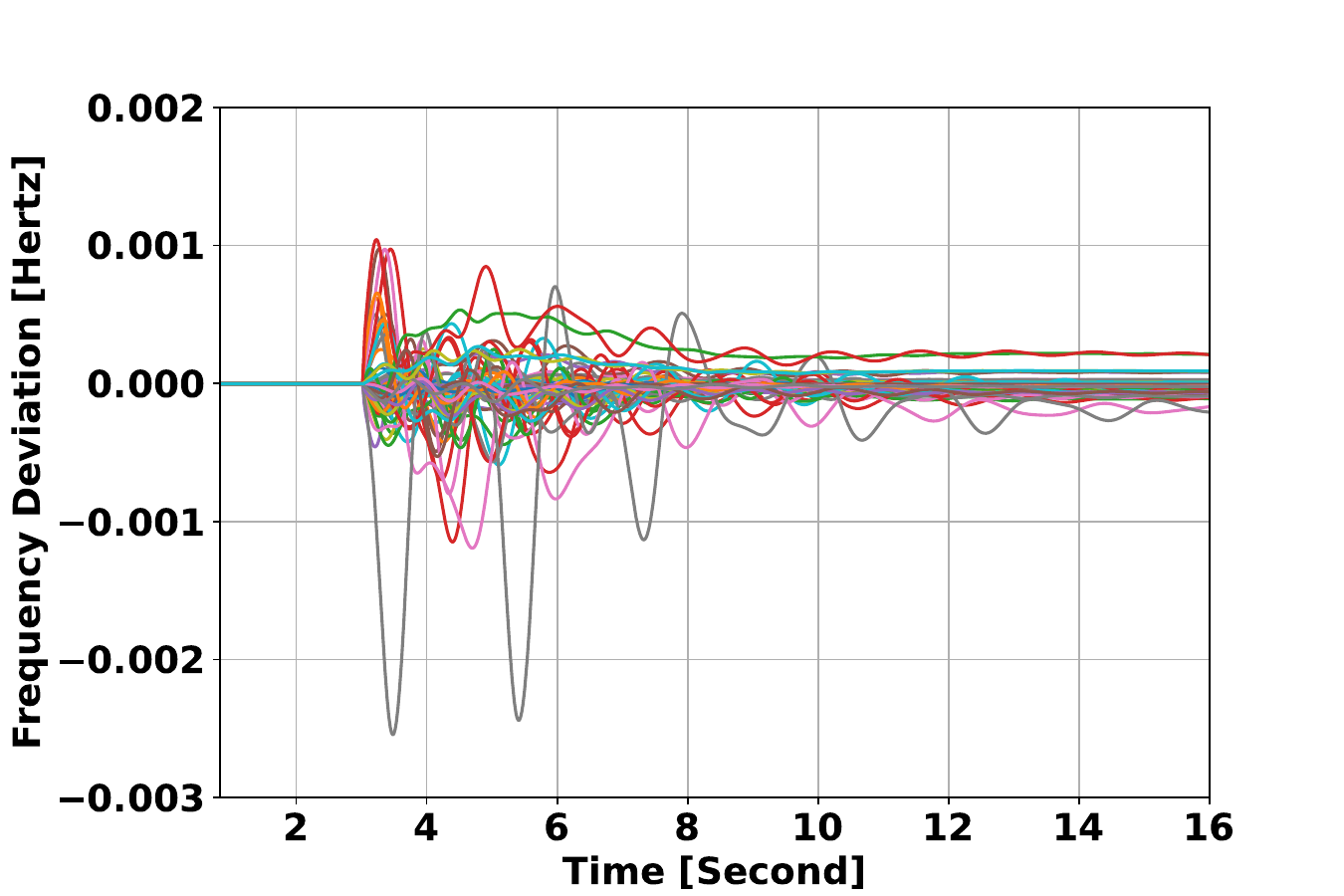}
    \label{fig:NoIes_voltage}
}
\caption{Voltage and frequency response of the data center load connected to the grid with co-located integrated SMR and BESS systems.}   \vspace{-15pt}

\label{withIes}
\end{figure}

An IES with a total capacity of 60 MW was then integrated with the data center, providing local generation support and fast-response capability through a combination of SMR baseload power and battery energy storage. In the grid-connected IES configuration, the transient responses of voltage and frequency for the same contingencies are presented in Fig.~\ref{withIes}.

Comparison of the results indicates that the IES significantly improves the stability of the data center interconnection. Specifically, the presence of the IES reduces voltage fluctuations, limits frequency deviations during disturbances, and provides faster recovery to pre-fault conditions.  
The integration of the IES enhances operational reliability to transient faults and ensures a stable power supply for sensitive data center operations. Results show that a data center with an IES not only supports its own load reliably but also contributes to the overall stability of the connected grid.

\section{Conclusion}\label{conc}
This study developed and analyzed a dynamic model of an Integrated Energy System (IES) that includes a Small Modular Reactor (SMR) and a Battery Energy Storage System (BESS) to support the reliable operation of grid-connected data centers. The SMR is a clean energy source that produces both electricity and thermal energy, making it ideal for data centers, which need both high computational power and continuous cooling. The SMR is modeled using a modified GE General Governor (GGOV1) framework with setpoint-based droop control, allowing it to respond to changes in electrical and thermal load. The BESS supports the SMR by providing fast dynamic support during transients through a proportional-integral control scheme, quickly stabilizing frequency and compensating for the SMR’s slower thermal response. 
The developed computational-thermal load model accounts for realistic variations in data center demand, accurately simulating its interaction with the power system. Simulations on the IEEE 118-bus test network showed that the IES-equipped data center had improved dynamic performance, including reduced voltage fluctuations, smaller frequency variations, and faster recovery after faults compared to conventional setups. These results demonstrate that coordinating nuclear and battery-based resources within an IES framework enhances both local reliability and overall system stability. 

Potential future work will focus on optimization-based scheduling, long-term economic analysis, and integrating digital twin technology for real-time monitoring, predictive control, and optimization of IES-equipped data centers. 
\textcolor{black}{Future work will also include a sensitivity analysis of BESS capacity to systematically evaluate its impact on system stability and resilience under varying data center load conditions.}

\bibliographystyle{IEEEtran}
\bibliography{ref}

\end{document}